\documentclass[conference, letterpaper]{IEEEtran}
\IEEEoverridecommandlockouts
\usepackage{cite}
\usepackage[left=2.55cm,right=2.55cm,top=1.9cm]{geometry}
\usepackage{amsmath,amssymb,amsfonts}
\usepackage{algorithmic}
\usepackage{graphicx}
\usepackage{textcomp}
\usepackage{xcolor}
\def\BibTeX{{\rm B\kern-.05em{\sc i\kern-.025em b}\kern-.08em
    T\kern-.1667em\lower.7ex\hbox{E}\kern-.125emX}}
\begin{document}

\title{ARSecure: A Novel End-to-End Encryption Messaging System Using Augmented Reality }

\author{
\IEEEauthorblockN{
Hamish Alsop \IEEEauthorrefmark{1},
Douglas Alsop\IEEEauthorrefmark{1},
Joseph Solomon \IEEEauthorrefmark{1},
Liam Aumento \IEEEauthorrefmark{1}, \\
Mark Butters \IEEEauthorrefmark{1},
Cameron Millar \IEEEauthorrefmark{1},
Yagmur Yigit\IEEEauthorrefmark{1}, 
Leandros Maglaras\IEEEauthorrefmark{1},
Naghmeh Moradpoor\IEEEauthorrefmark{1}}
\IEEEauthorblockA{
\IEEEauthorrefmark{1} School of Computing, Engineering and the Build Environment, \\ Edinburgh Napier University, United Kingdom \\
Email: 40342915@live.napier.ac.uk, 40531101@live.napier.ac.uk, 40528837@live.napier.ac.uk,  \\ 40533334@live.napier.ac.uk,  40582992@live.napier.ac.uk, 40583040@live.napier.ac.uk, \\ yagmur.yigit@napier.ac.uk, L.Maglaras@napier.ac.uk, n.moradpoor@napier.ac.uk}
}

\maketitle

\begin{abstract}
End-to-End Encryption (E2EE) ensures that only the intended recipient(s) can read messages. Popular instant messaging (IM) applications such as Signal, WhatsApp, Apple's iMessage, and Telegram claim to offer E2EE. However, client-side scanning (CSS) undermines these claims by scanning all messages, including text, images, audio, and video files, on both sending and receiving ends. Industry and government parties support CSS to combat harmful content such as child pornography, terrorism, and other illegal activities. In this paper, we introduce ARSecure, a novel end-to-end encryption messaging solution utilizing augmented reality glasses. ARSecure allows users to encrypt and decrypt their messages before they reach their phone devices, effectively countering the CSS technology in E2EE systems.
\end{abstract}


\section{Introduction}
\label{sec:intro}

Mobile devices are susceptible to physical over-the-shoulder eavesdropping and digital eavesdropping, often through malware. Social media and communication apps are mostly used, making up nearly half of global Internet traffic. However, downloaded apps may contain hidden stalkerware, a type of spyware that lets attackers monitor activity and access personal information. Stalkerware can record, monitor, or access email, social media, stored media, SMS, and chat apps, enabling Intimate Partner Stalking (IPS) where abusers e-follow current or former partners. Modern E2E encryption are failing to protect privacy, allowing malactors to perform spying activities. The use of desktop clients with shared system states that are open to compromise further questions the robustness of these systems \cite{chowdhury2023threat}.

As stated in many previous works, E2EE functionality can be compromised by recent technologies like client-side scanning (CSS), endpoint filtering, or local processing. A CSS uses an application that scans text, images, audio or video on a user's mobile phone in clear. If the CSS system finds a match from a specific list, it may prevent the user from sending the message or report it to law enforcement authorities\cite{scheffler2023sok}.  Moreover, European Police Chiefs are supporting CSS as a tool for ensuring public safety across social media platforms\footnote{https://www.europol.europa.eu/media-press/newsroom/news/european-police-chiefs-call-for-industry-and-governments-to-take-action-against-end-to-end-encryption-roll-out}.

The authors in \cite{abelson2024bugs} raise many questions related to the legitimate use of the technology such as: How is the list of targeted materials for client-side scanning compiled?  What privacy safeguards are implemented to prevent third parties from exploiting these scanning mechanisms to extract data? The paper concludes that it is uncertain whether CSS systems can be implemented securely in a way that ensures invasions of privacy are proportionate. 

In a recent work, the authors introduced the use of an encrypted keyboard to address the issues raised by the use of CSS. The encrypted keyboard  can be enabled by the user on his phone device and be used to encrypt or decrypt a message \cite{alatawi2022exploring}. 
The problem with this software solution is that it is still installed in the mobile phone of the user that is unsafe. 

Some earlier prototypes that were already built included a second mobile phone \cite{maglaras2022end} or Rasberry pi device \cite{velagala2022enhancing} that played the role of the secure device that encrypts and decrypts the messages before those are sent to the recipient through a normal messaging application. Here, we introduce ARSecure, a novel end-to-end encrypted messaging solution utilizing augmented reality glasses. ARSecure allows users to encrypt and decrypt their messages before they reach their phone devices, effectively countering the CSS technology in E2EE systems.

\section{Proposed ARSecure Mechanism}
\label{sec:proposed}

The proposed system combines encryption technologies with innovative technologies (augmented reality equipment) to decouple privacy preservation of messages from the ‘weak’ mobile device of the user. In order to achieve holistic end-to-end encryption, we use augmented reality smart glasses. These glasses can be used to form messages, encrypt and decrypt texts, and send the messages to other glasses users. We used the 2022 version of Vuzix Blade \footnote{https://www.vuzix.com/products/vuzix-blade-2-smart-glasses} that provides enhanced functionality for hands-free mode with two-way audio support (listening to voice commands and broadcasting sounds) providing an eyepiece computer with a transparent Waveguide display. 

To safeguard the privacy of users, all messages are encrypted using public/private cryptography models (e.g., Diffie-Hellman).  These keys can be generated through a central entity that is built for the project or by using PGP (Pretty Good Privacy) generators. 

\begin{figure}[htbp]
    \centering
    \includegraphics[width=\linewidth]{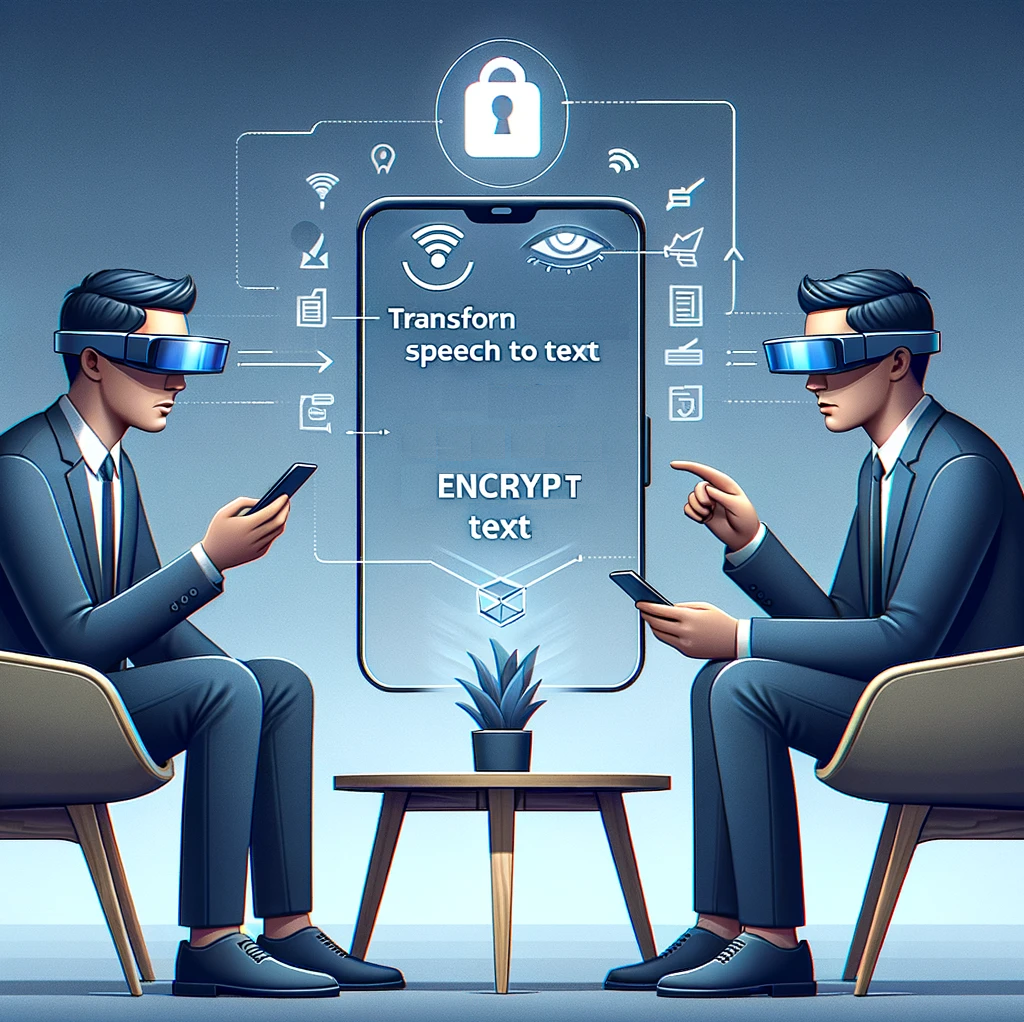}
    \caption{Illustration of ARSecure Mechanism}
    
    \label{BMFA}
\end{figure}

\subsection{Processes of ARSecure}
Below, we present the basic operations of ARSecure. 

{\bf Registration Module}
After the user has successfully booted the AR Glasses, he will be prompted to complete the user registration module. The registration of users will take place
on the AR Glasses once an internet connection is established.

{\bf Authentication Module}
The initial module to be implemented in the AR device  is the authentication module. Upon launching the chat application on any device, the authentication module is activated.  The only thing the authentication module will ask for is a password and not biometrics, PIN, draw pattern, face Identification, or voice Identification. Biometric authentication and Multi-Factor Authentication methods could be added for safer authentication of the user.
 
The {\bf Message creation creation and sending module}:

\begin{enumerate}
\item User talks to the smart glasses
\item Smart glasses convert speech to text
\item Text is encrypted using the private key of the user
\item Message is sent to the server and stored
\end{enumerate}

{\bf The receipt and digestion of a message module}:

\begin{enumerate}
\item Message is pulled from the server by the recipient's glasses
\item The message is decrypted on the smart glasses
\item The message is  shown on the end user
\end{enumerate}

{\bf End-to-End Encryption Methods}
The mechanism offers end to end encryption functionality using a separate secure hardware device. Not only does this method provide a solution to eavesdropping/Man in the Middle attacks, but could also help protect against CCS or keyloggers given the fact that the glasses use voice as input. 

The application is easily portable to any android operating system and is public through Github\footnote{https://github.com/TheFruggg/VuzixApp}.


\section{Conclusion}
\label{sec:conclusion}
In this article we present our novel ARSecure prototype. The current application uses augmented reality glasses for creating and digesting encrypted messages, in the future we will be able to replace it with a smart overlay screen or micro controller enabled device. This will make the application more user friendly while keeping the safe and secure characteristics. In terms of user authentication in future prototypes we may choose to add support for pin, face, or biometrics or voice recognition options.

\bibliographystyle{IEEEtran}
\bibliography{ref}

\end{document}